\documentclass[12pt]{article}

\usepackage{scicite}
\usepackage{times}
\usepackage{graphicx,amsmath,amssymb,color,soul} 

\newenvironment{sciabstract}{%
\begin{quote} \bf}
{\end{quote}}

\newcounter{lastnote}

\title{An accreting pulsar with extreme properties drives an ultraluminous x-ray source in NGC 5907}

\author
{Gian Luca Israel$^{1,\ast}$, Andrea Belfiore$^{2}$,  Luigi Stella$^{1}$,  Paolo Esposito$^{3,2}$, 
\\ Piergiorgio Casella$^{1}$, Andrea De Luca$^{2}$, Martino Marelli$^{2}$, 
\\ Alessandro Papitto$^{1}$,  Matteo Perri$^{4,1}$, Simonetta Puccetti$^{4,1}$, 
\\ Guillermo A. Rodr\'{\i}guez Castillo$^{1}$,  David Salvetti$^{2}$,  Andrea Tiengo$^{5,2}$,
\\ Luca Zampieri$^{6}$, Daniele D'Agostino$^{7}$, Jochen Greiner$^{8}$, Frank Haberl$^{8}$, 
\\ Giovanni Novara$^{5,2}$,  Ruben Salvaterra$^{2}$,  Roberto Turolla$^{9}$,
\\ Mike Watson$^{10}$, Joern Wilms$^{11}$, Anna Wolter$^{12}$\medskip\\
\normalsize{$^{1}$ Osservatorio Astronomico di Roma, INAF, via Frascati 33, I-00078 Monteporzio Catone, Italy}\\
\normalsize{$^{2}$ Istituto di Astrofisica Spaziale e Fisica Cosmica, INAF, via E. Bassini 15, I-20133 Milano, Italy}\\
\normalsize{$^{3}$ Anton Pannekoek Institute for Astronomy, University of Amsterdam,}\\
\normalsize{Postbus 94249, NL-1090GE Amsterdam, The Netherlands }\\
\normalsize{$^{4}$ ASI Science Data Center, via del Politecnico snc, I-00133 Roma, Italy}\\
\normalsize{$^{5}$ Scuola Universitaria Superiore IUSS Pavia, piazza della Vittoria 15, I-27100 Pavia, Italy}\\
\normalsize{$^{6}$ Osservatorio Astronomico di Padova, INAF, vicolo dell'Osservatorio 5, I-35122 Padova, Italy}\\
\normalsize{$^7$ Istituto di Matematica Applicata e Tecnologie Informatiche ``E. Magenes'', CNR,} \\
\normalsize{via de Marini 6, I-16149 Genova, Italy}\\
\normalsize{$^{8}$ Max-Planck-Institut f\"ur extraterrestrische Physik, Giessenbachstr., D-85748 Garching, Germany}\\
\normalsize{$^{9}$ Dipartimento di Fisica e Astronomia, Universit\`a di Padova, }\\
\normalsize{via F. Marzolo 8, I-35131 Padova, Italy}\\
\normalsize{$^{10}$ Department of Physics and Astronomy, University of Leicester, LE1 7RH Leicester, UK}\\
\normalsize{$^{11}$ Dr. Karl-Remeis-Sternwarte and Erlangen Centre for Astroparticle Physics,}\\
\normalsize{ Sternwartstr. 7, D-96049 Bamberg, Germany}\\
\normalsize{$^{12}$ Osservatorio Astronomico di Brera, INAF, via Brera 28, I-20121 Milano, Italy}\\\smallskip\\
\normalsize{$^\ast$E-mail:  gianluca@oa-roma.inaf.it}
}

\date{}

\def\xmm {\emph{XMM--Newton}}
\def\nustar  {\emph{NuSTAR}}
\def\cxo {\emph{Chandra}}
\def\swift {\emph{Swift}}

\def\hst {\emph{HST}}
\def\src {\mbox{NGC\,5907}}
\def\srcfull {\mbox{NGC\,5907 ULX}}

\def\lum {\mbox{erg s$^{-1}$}}

\begin {document} 

\baselineskip24pt

\maketitle 

\begin{sciabstract}
Ultraluminous x-ray sources (ULXs) in nearby galaxies shine brighter than any X-ray source in our Galaxy. 
ULXs are usually modeled as stellar-mass black holes (BHs) accreting at very high rates or intermediate-mass 
BHs. 
We present observations showing that \srcfull\ is instead an x-ray accreting neutron star 
(NS) with a spin period evolving 
from 1.43~s in 2003 to 1.13~s in 2014. It has an isotropic peak luminosity of $\sim$1000 times  the 
Eddington limit  for a NS  at 17.1~Mpc. Standard 
accretion models fail to explain its luminosity, even assuming beamed emission, but a strong multipolar 
magnetic field can describe its properties. These findings suggest that other extreme ULXs (x-ray luminosity 
$\geq$10$^{41}$ \lum) might harbor NSs.
\end{sciabstract}

Ultraluminous x-ray sources (ULXs) are observed in off-nucleus regions of nearby galaxies and have x-ray 
luminosities 
in excess of a few $10^{39}$~\lum, which is the Eddington luminosity  ($L_{\mathrm{Edd}}$) for a black hole 
(BH) of 10 $M_\odot$ \cite{fabbiano06}. 
The $L_{\mathrm{Edd}}$ sets an upper limit on the accretion luminosity 
($L_{\mathrm{acc}}$) of a compact object steadily accreting, since for $L_{\mathrm{acc}}> L_{\mathrm{Edd}}$ 
accretion will be halted by radiation forces. For spherical accretion of fully ionized hydrogen, the  
limit can be written as $L_{\mathrm{Edd}}=4\pi cG M m_{\mathrm{p}}/\sigma_{\mathrm{T}}\simeq1.3\times10^{38} 
(M/M_{\odot})$~\lum, where $\sigma_{\mathrm{T}}$ is the Thomson scattering cross section, $m_{\mathrm{p}}$ 
is the proton mass, and $M/M_{\odot}$ is the compact object mass in solar masses; for a 
1.4\,$M_{\odot}$ neutron star (NS), the maximum accreting luminosity is $\sim$$2\times10^{38}$~\lum.
The high luminosity of ULXs has thus been explained as accretion 
at or  above the Eddington luminosity onto BHs of stellar origin 
($<$80--100~$M_\odot$), or onto intermediate-mass ($10^3$--$10^5$~$M_\odot$) BHs \cite{poutanen07,zampieri09}.
However, if the emission of ULXs were beamed over a fraction $b<1$ of the sky, their true luminosity, and thus 
also the compact object mass required not to exceed $L_{\mathrm{Edd}}$, would be reduced by the same factor. 
This possibility, together with the recent identification of two accreting NSs associated with the 
$\sim$10$^{40}$~\lum\ M82 X-2 \cite{bachetti14} and NGC\,7793 P13 \cite{israel16a,fuerst16} x-ray sources, 
have brought support to the view that most low-luminosity 
ULXs likely host a NS \cite{fragos15} or a stellar-mass BH \cite{motch14}. For the most extreme ULXs with x-ray 
luminosity exceeding a few $\times 10^{40}$~\lum, BHs with masses in excess of 100\,$M_{\odot}$ are still commonly 
considered\cite{sutton12,mushtukov15}.
Despite several searches for coherent x-ray pulsations,
no other ultraluminous x-ray source has been found to host a NS so far \cite{doroshenko15}.\\

Within the framework of ``Exploring the X-ray Transient and variable Sky'', EXTraS \cite{deluca16}, a project 
aimed at characterizing the variability of x-ray sources observed with the X-ray Multi-Mirror Mission (\xmm)
satellite, we have undertaken a systematic search for new x-ray pulsators in the archival data of the European 
Photon Imaging Camera (EPIC) instrument \cite{struder01,turner01}. During the analysis of a $\sim$40\,ks-long 
observation carried out from  9 to 10 July 2014 [see table\,S1 in \cite{sciencemm}] and pointed to the edge-on spiral galaxy NGC\,5907 
at a distance of about 17.1\,Mpc \cite{tully13}, we found a coherent signal in the x-ray emission of 
\srcfull.  In fact, a prominent peak at a frequency of $\sim$0.88~Hz (5.8$\sigma$ detection 
significance),  corresponding to a period $\sim$1.137~s, was detected in the Fourier power spectrum of the 
0.2--12\,keV light curve of the source. A first period derivative, $\dot{P} \sim -5 \times $10$^{-9}$~s~s$^{-1}$, 
was also detected.  
A refined search, including a correction for the $\dot{P}$ term, allowed us to detect the signal in an earlier  
\xmm\ observation taken on February 2003, and in two observations carried out with the Nuclear Spectroscopic 
Telescope Array (\nustar) mission \cite{harrison13} taken in July 2014, 
with periods of $\sim$1.428~s in 2003, and  $\sim$1.136~s in 2014 (see Fig.\,1 and Table 1). 
In all  cases, a strong first period derivative term is present (see Table\,1). The pulse shape is nearly 
sinusoidal, while the pulsed fraction (the semi-amplitude of the sinusoid divided by the average count rate) 
is energy dependent and increases from about 12\% at low energies ($<$2.5~keV) to $\sim$20\% in the hard band 
($>$7~keV; Fig.\,1). 

To  derive constraints on the orbital period ($P_{\mathrm{orb}}$), we applied a likelihood analysis to the 
two 2014 \nustar\ observations (see supplementary online text), which have the longest baseline. By assuming 
a circular orbit (as in the case of M82 X-2), a most-probable $P_{\mathrm{orb}} = 5.3^{+2.0}_{-0.9}$~d 
(see Fig.\,2, where 1, 2 and 3$\sigma$ confidence levels are shown) is inferred, with a projected semi-axis $a \sin{i}= 2.5^{+4.3}_{-0.8}$~light-s 
(about 7.5$\times10^{10}$~cm), where $i$ is the orbital inclination. 
Though we cannot exclude orbits with period of the order of one month or longer based on the timing analysis 
alone, we noticed that periods longer then 20~d would require a very high-mass companion ($M>100~M_{\odot}$) and are therefore unlikely (see Fig.\,2).

In the subsequent analysis, we assume an  average spin-up rate of  
$-(8.1\pm0.1)\times10^{-10}$~s~s$^{-1}$ implying a spin-up timescale $P/\dot P \sim 40$~yr \cite{lipunov82}. 
This value is derived from the ratio $\frac{\Delta P}{(T_{14}-T_{03})}$ where 
$\Delta P$ is the difference between the periods measured in February 2003 ($T_{03}$) and in July 2014 ($T_{14}$; 
see Table\,1). As this was obtained from a long baseline, it is largely unaffected by the orbital 
Doppler shift (which is instead present in each single dataset) and, therefore, can be considered a good estimate 
of the long-term average $\dot P$.

The luminosity measured with \swift, \cxo, \xmm, and \nustar\ (Fig.\,S2) display a pronounced variability 
both on long and short timescales \cite{sutton13,walton15}. 
The source was detected in $\sim$85\% of the observations, its bolometric luminosity \cite{sciencemm} ranged between  
$(2.6\pm0.3) \times 10^{40}$ and $(2.2\pm0.3)\times10^{41}$~\lum\ (by a factor of about 8). In late 
2013, \cxo\ and \xmm\ provided 3$\sigma$ upper 
limits on the luminosity of $\sim$$3\times 10^{39}$ and $\sim$$4\times10^{38}$~\lum, respectively (see Fig.\,S2 and 
table\,S1).

If isotropically emitted, the maximum luminosity of \srcfull\ is $\sim$1000 times the
Eddington limit for a 1.4-$M_{\odot}$ NS; it
thus challenges current models for magnetospheric accretion.
NSs can attain highly super-Eddington
luminosities only if their magnetic fields is very high: a luminosity of  $\sim$$1000\ L_{\mathrm{Edd}}$
would require a field strength of $>$$10^{15}$~G \cite{mushtukov15,dallosso15}.
However, for such a field and the mass inflow rate
required to sustain the observed luminosity, the $\sim$1~s rotation of the NS and its
magnetosphere would drag matter at the magnetospheric boundary (a distance corresponding to 
the magnetospheric radius $r_{\mathrm{m}}$) so fast
that centrifugal forces exceeds gravity and accretion onto the NS surface is inhibited by the
so-called propeller mechanism \cite{illarionov75,stella86,sciencemm}.

If the emission of \srcfull\ were beamed over a fraction $b <1$ of the sky, then the
true source luminosity generated by accretion would be lowered by the same factor.
A solution is in principle possible for $b \sim 1/100$ and a surface field of
$\sim$$9\times 10^{12} $~G, as for such value the maximum luminosity the NS can attain
would be compatible with the observations and accretion onto the NS
surface could take place, without being inhibited by the propeller.
However, the mass accretion rates predicted by this solution would be so low
that the corresponding torques would be unable to spin-up the NS at the observed secular
rate. Therefore, we conclude that current magnetospheric
accretion models which are based on the assumption that the NS magnetic field is purely
dipolar are unable to explain the source properties.

A model capable of interpreting the properties
of \srcfull\ involves the presence of a multipolar magnetic field at the NS
surface of $B_{\mathrm{multi}} \sim (0.7$--$3) \times 10^{14}$~G, of which only $\sim(0.2$--$3) 
\times 10^{13}$~G is 
in the dipole component, and a moderate beaming from
$b\sim 1/25$ to $1/7$ [see \cite{sciencemm} and Fig.\,3 for details]. With such model, all conditions
required for the NS to accrete and generate the factor of $\sim$8 of observed true accretion 
luminosities and spin-up rate would be met.
Such a magnetic field configuration is similar to that envisaged for magnetically-powered
neutron star, the so-called magnetars \cite{thompson95,tiengo13}.

The transient x-ray pulsar we detected in \srcfull\ demonstrates that accreting NSs can achieve extreme
luminosities not foreseen in current accretion models. Such high luminosities are often displayed by many
ULXs which have previously been classified as accreting black holes.
A multi-component strong magnetic field is necessary to account for the properties of \srcfull.

\clearpage

\bibliography{bibliosci}
\bibliographystyle{Science}

\clearpage

\noindent {\bf \large \sc acknowledgments\,} \\
EXTraS is funded from the EU's Seventh Framework Programme under grant agreement no. 607452. This research 
is based on observations obtained with \xmm, an ESA science mission with instruments and contributions 
directly funded by ESA Member States and NASA. This work also made use of data from \nustar, a mission led by 
Caltech, managed by the JPL, and funded by NASA, and from \swift, which is a NASA mission with participation 
of the Italian Space Agency and the UK Space Agency. GLI, PC, LZ and AW  acknowledges funding from the ASI--INAF
 contract NuSTAR I/037/12/0. PE acknowledges funding in the framework of the NWO Vidi award A.2320.0076. 
AP acknowledges funding from the EU's Horizon 2020 Framework Programme for Research and Innovation under the Marie 
Sk\l{}odowska-Curie grant agreement 660657-TMSP-H2020-MSCA-IF-2014. GLI is grateful to Thomas Tauris for 
useful comments on an earlier version of the manuscript; PE thanks Alexander Mushtukov for interesting discussions. 
The EXTraS project acknowledges the usage of computing facilities at INAF's Astronomical Observatory of Catania 
and Astronomical Observatory of Trieste.
The data presented here can be found in the supplementary materials; raw x-ray observations
can be retrieved through the ESA/\xmm\ archive interface (http://www.cosmos.esa.int/web/xmm-newton/xsa) and 
the NASA's ones for \cxo\ (http://cda.harvard.edu /chaser/) and \swift\ (http://heasarc.gsfc.nasa.gov/cgi-bin/W3Browse/swift.pl); 
see Table\,S1 for the identification number of the observations. 


\clearpage
\begin{table}
\centering
  \noindent  {\bf Table\,1  Timing properties of the \srcfull\ pulsar.} ~ 1$\sigma$ confidence level 
  is assumed for the uncertainties. Observational dates are expressed in modified Julian 
  days (MJD).\label{tabtime}\\
\begin{tabular}{lrrrr}
 \\
\hline
Start Date &2003 Feb 28& 2014 Jul 09& 2014 Jul 09& 2014 Jul 12\\
Mission &\xmm & \nustar &\xmm & \nustar\ \\
Epoch (MJD) & 52690.9&  56848.0& 56848.2& 56851.5\\
$P$ (s) &1.427579(5) & 1.137403(1)&1.137316(3) & 1.136042(1)\\
$\dot{P}$ ($10^{-9}$ s s$^{-1}$) &--9.6(9) & --5.2(1)& --5.0(5)& -4.7(1)\\
\hline \\
\end{tabular}
\end{table}

\begin{figure}
\resizebox{\hsize}{!}{\includegraphics{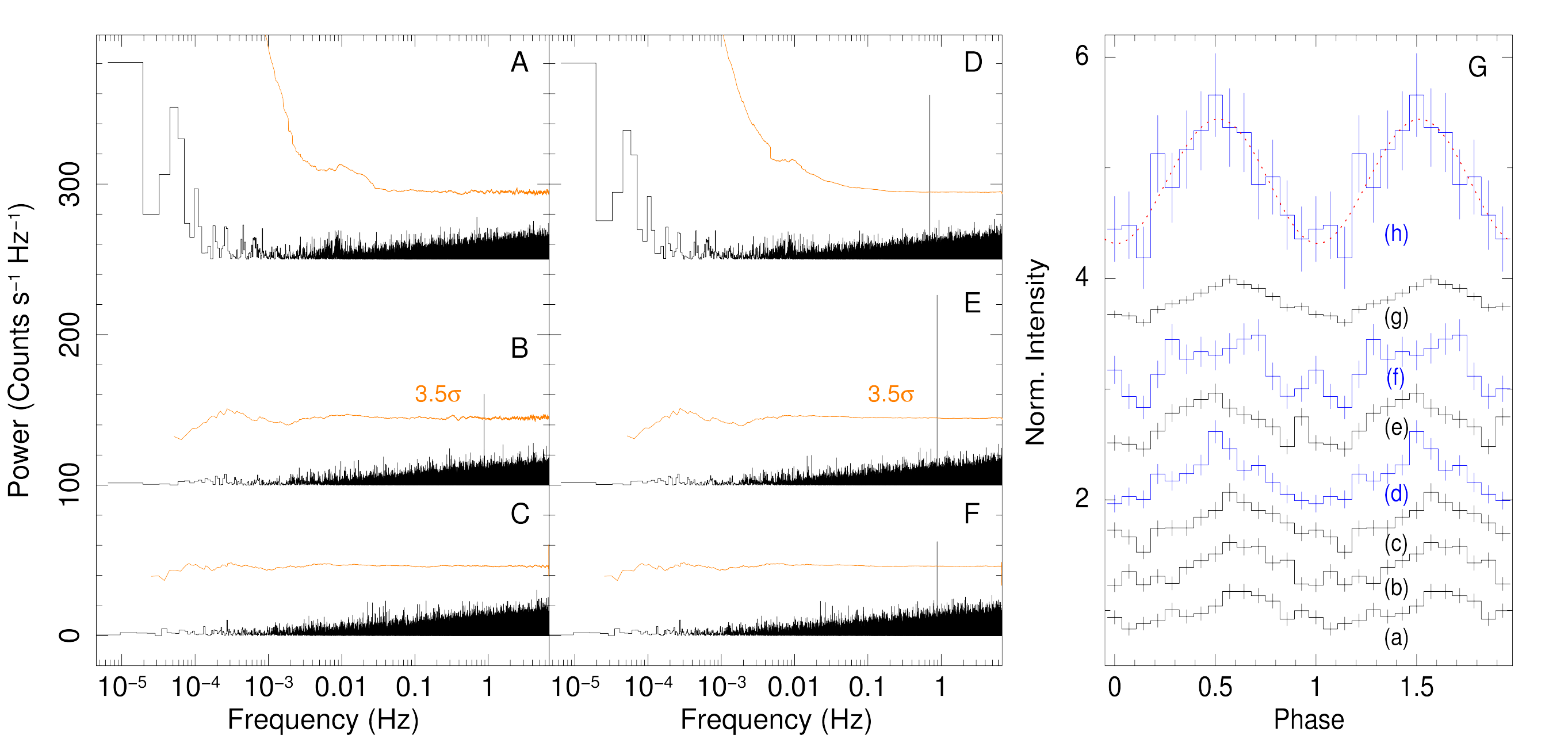}}
\noindent {\bf Fig.\,1. Detection and study of the pulsations observed in the extreme ULX in NGC\,5907.} ~ 
Arbitrarily  shifted (along the y-axis) power spectral density (PSD) of the 
0.2--12~keV (\xmm) and 3--30~keV (\nustar) \srcfull\ light curves of three out of the four datasets where 
pulsations have been detected: \xmm\ observations  of 2003-20-02 (A) and 2014-07-09/10 (B), and \nustar\ 
observation of 2014-07-09/10 (C). The calculated 3.5$\sigma$ detection 
threshold \cite{israel96} is shown for each PSD (light orange solid lines). In the central panel, we show 
the PSD of the same light curves (A, B and C) after correcting the photon arrival times for the $\dot{P}$ 
term (D, E and F).
The 2014 \xmm\  (black solid lines) and \nustar\ (blue solid lines) simultaneous pulse profiles are shown 
in the right panel (G). From the bottom to the top, energy intervals are: (a) 0.2--2.5~keV, (b) 2.5--4~keV, 
(c) and (d) 4--7~keV, (e) and (f) 7--12~keV, (g) 0.2--12~keV, and (h) 3--30~keV. Profiles are arbitrarily 
shifted along the y-axis and two cycles are shown for clarity.
\end{figure}
\clearpage 

\begin{figure}
\resizebox{\hsize}{!}{\includegraphics{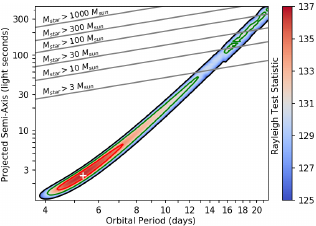}}
\noindent {\bf Fig.\,2. Orbital period constraints from the likelihood analysis of the 2014 x-ray datasets.} 
~ Constraints on the orbital parameters of \srcfull\ obtained with a direct 
likelihood analysis of the 2014 \nustar\ observations. For each point in a grid of projected semi-axis $a$ 
and orbital period $P_{\mathrm{orb}}$, we show the maximum Rayleigh test statistic $R$ \cite{sciencemm}.  This is 
obtained scanning over the spin parameters, period $P$ and its time derivative $\dot{P}$, and the epoch of 
ascending nodes, $T_0$ as described in \cite{sciencemm}. 
The best-fit values of $a$ and $P_{\mathrm{orb}}$ are indicated with a cross. The contours mark the $R$ confidence 
levels of 1, 2, and 3$\sigma$. 
The solid parallel lines indicate particular configurations for which the orbital inclination
and the masses of the two objects are held fixed.
According to Kepler's third law, the mass of the companion to a 1.4 $M_{\odot}$ neutron
star is indicated for a system observed edge-on. 
As the inclination of the system is unknown, these values must be taken as lower bounds
to the actual mass of the companion star.
\end{figure}

\clearpage 

\begin{figure}
\resizebox{\hsize}{!}{\includegraphics{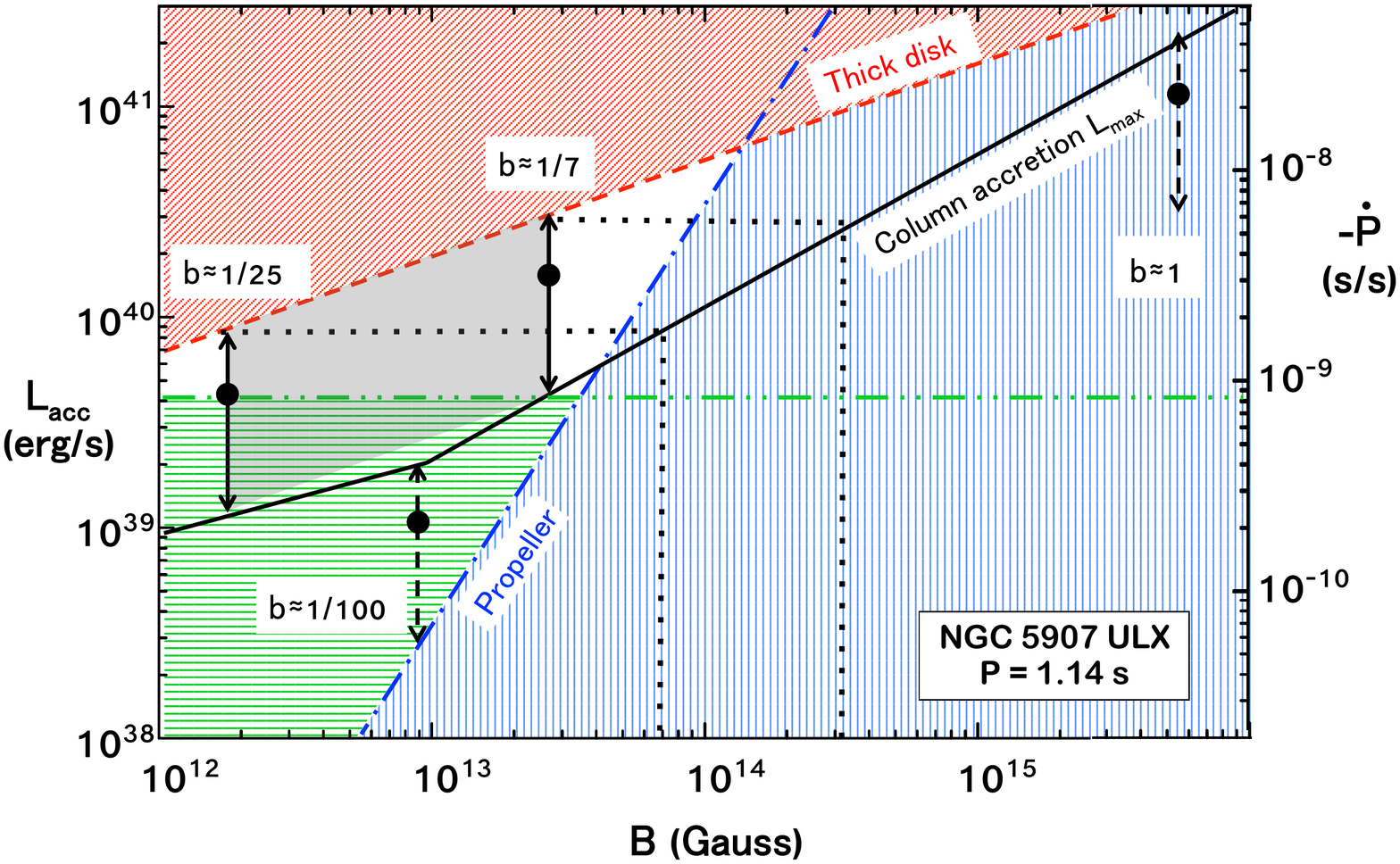}}
\noindent {\bf Fig.\,3. Accretion luminosity versus surface magnetic field contraints for \srcfull.} ~
The black solid line represents the maximum luminosity attainable via magnetic column-accretion
onto the NS\cite{mushtukov15}. Above the dashed (red) line, the energy released in the accretion disk 
down to the magnetospheric boundary exceeds the Eddington luminosity and the magnetospheric boundary 
is engulfed with the inflowing matter\cite{mushtukov15}. Below the dot-dashed (blue) line, the centrifugal 
drag by the rotation of the NS magnetosphere exceeds gravity and the propeller regime ensues, where little 
(if any) accretion takes place\cite{stella86}.
The double-dotted dashed (green) line represents the secular $\dot P$ measured for \srcfull\
(see the relevant Y-axis scale on the right); it is plotted here for the minimum
accretion luminosity that is required to give rise to it. Below this line the accretion rate would
not be sufficient to spin-up the NS at the observed rate. 
Double-arrowed segments represent the factor of $\sim$8 flux variation over which the source was detected. 
They are shown for different values of the beaming factor (and are correspondingly shifted in accretion 
luminosity by $b^{-1} = L_{\mathrm{iso}}/L_{\mathrm{acc}}$): dashed arrows represent solutions which
are not viable. A value of $b\sim 1/25 $--$1/7$ can reproduce all source features \cite{sciencemm},
for a magnetic dipole surface field of $\sim$$(0.2$--$3)\times 10^{13}$~G and a multipolar surface 
field of $\sim (0.7$--$3) \times 10^{14}$~G. 
\end{figure}
\clearpage

\thispagestyle{empty}

\begin{center}
  {\Large Supplementary Materials for} \\ \vspace{4mm}

{\large \bf An accreting pulsar with extreme properties drives an ultraluminous \\ \vspace{-3mm} 
  X-ray source in NGC 5907}\\ \vspace{-3mm}
{Gian Luca Israel,$^{\ast}$ Andrea Belfiore,  Luigi Stella,  Paolo Esposito, Piergiorgio Casella,
\\  \vspace{-4mm}   Andrea De Luca, Martino Marelli, Alessandro Papitto,  Matteo Perri,
\\ \vspace{-4mm}  Simonetta Puccetti, Guillermo A. Rodr\'{\i}guez Castillo,  David Salvetti,  Andrea Tiengo,
\\ \vspace{-4mm}    Luca Zampieri, Daniele D'Agostino, Jochen Greiner, Frank Haberl, 
\\ \vspace{-4mm}   Giovanni Novara,  Ruben Salvaterra,  Roberto Turolla,
\\ \vspace{-4mm}   Mike Watson, Joern Wilms, Anna Wolter}\\

{\small $^\ast$ Corresponding author. Email:  gianluca@oa-roma.inaf.it}
\end{center}

{\small {\bf This PDF file includes}:\\ \vspace{-3mm}
  \indent\indent\indent Materials and Methods\\  \vspace{-3mm}
  \indent\indent\indent Figs S1, S2, S3, S4\\ \vspace{-3mm}
  \indent\indent\indent Table S1 \\ \vspace{-3mm}
  \indent\indent\indent References (27-52)\\ \vspace{-3mm}
}

\clearpage
\setcounter{page}{2}

\noindent

\noindent
{\Large\bf Materials and Methods}

\noindent
{\large\bf Observations and data reduction }

\noindent We made use of seven \xmm, two \cxo,  five \nustar, and 151 \swift\ observations; see Table S1 for dates and 
exposures.\\ 

\noindent
{\bf \xmm} 

\noindent In  the \xmm\ observations, the positive--negative junction (pn) CCD camera \cite{struder01} of the EPIC 
instrument was operated in Full Frame mode, with a time resolution of 73.4 ms. The two metal oxide semi-conductor 
(MOS) 
CCD detectors \cite{turner01} were set in Full Frame mode, with 2.6 s time resolution. 
The data reduction was performed with the \xmm\ Science Analysis Software (SAS) v14.0 \cite{sas}. 
We restricted the analysis to the `good time' intervals (GTIs) for the relevant CCDs provided in the Processing 
Pipeline Subsystem (PPS) 
products available for 
each observation; this includes the screening out of time periods with flaring particle background. After a careful 
study of the brightness profile, we chose to extract the source spectra and event lists from 30$''$-radius circular 
regions around the \cxo\ source position, $\rm RA= 15^h15^m58.\!\!^{\rm s}62$ and $\rm Decl.=+56^\circ18'10.\!\!'3$ 
\cite{sutton13}, in the energy ranges of 0.3--10\,keV and 0.2--12\,keV, respectively 
(see Fig.\,S1); the background was estimated locally for each observation.
For the timing analysis, the arrival times of the source photons were shifted to the barycentre of the solar system 
using the SAS tool barycen and assuming the solar system ephemeris DE200 \cite{de200}.\\

\noindent
{\bf \nustar}

\noindent Each of the two \nustar\ telescopes has a focal plane module (FPM), consisting of a solid-state CdZnTe pixel 
detector surrounded by an anti-coincidence shield \cite{harrison13}. The two modules, FPMA and FPMB, are sensitive to 
photons in the 3--78~keV range and event times are recorded with 2-$\mu$s accuracy.
The raw event files of the observations (Table S1) were processed using the \nustar\ Data Analysis Software  
(NuSTARDAS, v1.6.0) which is part of the ``General and mission-specific tools to manipulate FITS files'' (FTOOLS)
package \cite{ftools}.
Calibrated and cleaned event files were produced using the nupipeline task with standard filtering criteria and 
the calibration files in the CALDB database (release of 2016-03-15). We used the nuproducts task to extract for 
each FPM the source and background energy spectra and the barycentre-corrected light-curves in the total range 
(3--78~keV) and in five energy bands (3--4, 4--5, 5--7, 7--12 and 12--30~keV).
We used circular apertures of radius 49$''$ (corresponding to $\sim$65\% of the encircled energy; see Fig.\,S1) 
centred on the \cxo\ source position. Background energy  spectra and light-curves were extracted using 
source-free circular regions with radius 98$''$ located on the same detector as the source. \\

\noindent {\bf \cxo\ }

\noindent All \cxo\  observations we used were carried out with the Advanced CCD Imaging Spectrometer 
(Spectroscopic array, ACIS-S) in full-imaging mode with time resolution of 3.2~s  \cite{garmire03}. To extract 
the spectra in the 0.2--10\,keV energy range, we used the specextract tool of the \cxo\ Interactive Analysis of 
Observation (CIAO, v4.8) software package \cite{ciao}. We used a circular region of 2$''$ to extract the source counts (see Fig.\,S1), 
and the background was estimated locally for each observation. Due to the contiguity of the two observations 
and the low statistics of the source counts, we combined the two spectra using combine\_spectra, which also 
averages the response matrices.\\

\noindent {\bf \swift}

\noindent 
The X-Ray Telescope (XRT) on board \swift\ uses a CCD detector sensitive to photons with energies between 0.2 and 
10 keV \cite{burrows05}. All observations used for this work were performed in imaging photon counting (PC) mode, 
which provides a time resolution of 2.507~s. Data were processed following standard procedures using 
FTOOLS \cite{ftools}. 
We extracted the source events from a circular region with radius of 30$''$ centred on the \cxo\ position of \src, 
while to evaluate 
the background we extracted events from a source-free region of 130$''$-radius, avoiding the plane of \src. 
The ancillary response files were generated with xrtmkarf and accounts for different extraction regions, 
vignetting and point-spread function corrections. We used the latest available spectral redistribution matrix 
(v014).\\

\noindent 
{\large\bf  Timing analysis}

\noindent  In the context of the EXTraS project, all the \srcfull\ event lists were screened through a blind 
search for periodic signals in an automatic fashion following the recipe outlined in\cite{israel96}.
We found a significant periodic signal from \srcfull\ only in the \xmm\ pn data of the last observation, 
performed in 2014. 
Then, we searched again for pulsations all the  \xmm\ and \nustar\ observations,
accounting for a strong time derivative of the spin frequency, and
found significant pulsations in 4 cases. We run through
a grid in $-2\times10^{-8}\,\mbox{s}^{-1}\leq\dot{\nu}/\nu\leq2\times10^{-8}\,\mbox{s}^{-1}$,
where $\nu$ is the spin frequency and $\dot{\nu}$ its first time derivative,
by stretching the time series with this transform: $t'=t+\dot{\nu}/\nu\times t^{2}$, where $t$ is the 
raw time and $t'$ the transformed one. In this way, a signal 
with a significant linear trend in frequency
is converted into a periodic signal: $\nu\times t'=\left(\nu+\dot{\nu}\times t\right)\times t$.
We run on each stretched time series a fast Fourier transform, with $2^{16}$ bins,
up to 10 \mbox{Hz}. In only 4 cases out of 7 we detected a peak above the
$3\sigma$ threshold, trials-corrected; in all four cases the
signal was well above the $10\sigma$ confidence level. We refined
the timing solutions through a direct likelihood approach (see below). 
We validated our findings by folding the 
events with PRESTO\cite{ransom02} at
the detection ephemeris. In 2003 the signal is clearly detected
by \xmm\ on February 20 but it is absent only 7
days afterwards, when the source was again observed by \xmm, in a very
similar observation, at the same count rate as before. 
In the latter observation, we inferred a 3$\sigma$ upper limit on the pulsed fraction of $\sim$12\%. \\ 

\noindent {\bf Direct Likelihood Analysis}

\noindent We refined the ephemeris of single detections and constrained the
orbital parameters using a  timing technique that relies
on unbinned likelihood analysis\cite{bai92,cowan11}. We assume a sinusoidal model for
the light curve of the source, where the spin frequency depends on several
parameters. When studying individual detections, these parameters are
only the spin frequency and its time derivatives. When studying the
orbit of the system, besides the spin frequency and its first time
derivative, we also have to account for the orbital parameters. In
both cases the probability density of detecting a photon at any rotational
phase $\theta\left(\overrightarrow{M},t\right)$, where $\overrightarrow{M}$
is the vector of parameters in the model, can be stated as: 
\begin{equation}\tag{S1}
P\left(\theta\right)=\frac{1}{2\pi}\times\left(1+A\times\cos\left(\theta-\phi\right)\right)
\end{equation}
where $0\leq A\leq1$ is the pulsed fraction and $\phi$ the spin period phase. In principle, the most
likely value of $\left(\overrightarrow{M},A,\phi\right)$ should be
computed by multiplying the likelihood of detecting each photon at
the time it was detected, given the set of parameters $\left(\overrightarrow{M},A,\phi\right)$, 
and by scanning the parameter space. The maximum
log-likelihood with respect to $A$ and $\phi$, under the hypothesis
that the signal is sinusoidally modulated, approaches the Rayleigh
test-statistic for small $A$. The Rayleigh test-statistic, that we define as 
\begin{equation}\tag{S2}
R=\frac{2}{N}\times\left[\left(\sum_{i=1}^{N}\sin\theta_{i}\right)^{2}+\left(\sum_{i=1}^{N}\cos\theta_{i}\right)^{2}\right] 
\end{equation}
where 
$\theta_{i}=\theta\left(\overrightarrow{M},t_{i}\right)$
is the phase expected by our model at the time $t_{i}$ associated
to the $i$-th photon, out of $N$, is much faster to compute than
the likelihood itself. Therefore, this result allows us to replace
the likelihood with $R$, when searching for the most likely solution
and also for studying the likelihood profiles associated to our timing
model.
In particular, we can consider two nested models, one obtained from
the other by freezing $k$ of its $m$ parameters: we can try to reject
the nested model by comparing the maxima of their log-likelihood.
The increase in log-likelihood
produced by further $k$ free parameters  follows a $\chi^{2}$ distribution
with $k$ degrees of freedom (dof): $2\times\Delta\log\mathcal{L}_{k}\sim\chi_{k}^{2}$.
By replacing the log-likelihood with $R$, we can determine a region
in a $k$-dimensional space where the P-value associated to a specific
drop in $R$ meets our desired confidence level. In particular, the $P$ and $\dot{P}$ 
uncertainties  are obtained through the likelihood analysis, varying each parameter 
independently. The epochs were chosen close to the center of each observation 
in order to minimize the correlation between $P$ and $\dot{P}$.\\

\noindent {\bf Constraining the Orbital Parameters}

\noindent We choose the two \nustar\ observations to constrain the orbital parameters
of the system, through a direct likelihood analysis, in order to obtain confidence
regions and a relation between the orbital period $P_{\mathrm{orb}}$ and its projected
semi-axis $a$. We used the two \nustar\ observations
because they are longer and more closely spaced than the \xmm\ observations,
yet a change in $\dot{\nu}$ is apparent, likely resulting from the orbital modulation. 
We consider a simple model of a circular
binary orbit, in order to limit the number of free parameters while
still producing some useful constraints. The time delay due to the
orbital motion can be written as 
\begin{equation}\tag{S3}
\left(\Delta t\right)_{\mathrm{orb}}=a\times\sin\left(\frac{2\pi}{P_{\mathrm{orb}}}\times\left(t-T_0\right)\right)\label{eq:orb_dt} 
\end{equation}
 where $T_0$ is the epoch of the ascending node. From this relation,
we can correct the event times, accounting  for the orbital
motion of the system  and intrinsic  spin frequency derivative $\dot{\nu}$. 
We can then find the most likely spin frequency
$\nu$, pulsed fraction $A$, and spin phase $\phi$, by maximizing
the Rayleigh test-statistic $R$. By scanning over the 5 parameters
(2 rotational, and 3 orbital) 
and fixing $a$ and $P_{\mathrm{orb}}$
over a grid, we maximize $R$ over the other 3 parameters at each
point in the grid and obtain Fig.\,2. Contour
levels obtained by using the prescription in the previous paragraph, 
mark changes of {2.28, 6.13, 11.73} in $R$,
which correspond to confidence levels of 1, 2, and 3$\sigma$, respectively,
for a $\chi^{2}$ distribution with 2 dof {in
$2\times\Delta\log\mathcal{L}_{k}$}. 
Overlaid in the same plot are the lines indicating the maximum allowed mass
for a companion to a 1.4-$M_{\odot}$ NS, compatible with $a$ and $P_{\mathrm{orb}}$. While we cannot exclude 
from likelihood profiles alone an arbitrarily large value of $a$ and
$P_{\mathrm{orb}}$, we take 30 $M_\odot$ as indicative value for the mass of the companion. Correspondingly,  
we can estimate an upper limit on the variation of the frequency $\nu$ due to the Doppler modulation, 
at 3$\sigma$ as: 
\begin{equation}\tag{S4}
 \left|\left(\Delta \nu\right)_{\mathrm{orb}}\right|=\frac{2\pi}{P_{\mathrm{orb}}}\times a\times 
\nu <7.3\times10^{-4}\mbox{Hz}.
\end{equation}
This implies that the difference in $\nu$ between 2003 and 2014 is 
essentially all due to a change in the intrinsic spin period of the pulsar, while the
orbital modulation is a secondary effect. Larger masses implies smaller contributions  of 
the orbit to $\nu$. In the same way we can derive from the above relation for $\left(\Delta t\right)_{\mathrm{orb}}$, 
an estimate of the amplitude of the time derivative of the spin frequency $\dot{\nu}$ as 
\begin{equation}\tag{S5}
\left|\left(\Delta\dot{\nu}\right)_{\mathrm{orb}}\right|=\left(\frac{2\pi}{P_{\mathrm{orb}}}\right)^{2}\times 
a\times \nu<3.1\times10^{-9} \mbox{Hz} s^{-1}.
\end{equation}
This implies that the discrepancy between the secular spin-up and
the instantaneous spin-up measured in the single observations cannot
fully be accounted for by the orbital modulation alone.
\\

\noindent
{\large\bf Phase Averaged and Pulse Phase Spectroscopy}

\noindent We fit simultaneously the \xmm, \cxo\ and \nustar\ spectra using XSPEC v.12.9.0 \cite{arnaud96}. For all 
tested models, we added a multiplicative factor to account for the different normalizations and 
uncertainties in the inter-calibration of the various instruments. We grouped \xmm\ spectra so as 
to have at least 100 counts per energy bin and \cxo\ and \nustar\ ones with a minimum of 50 counts 
per bin. We analyzed only data in the 0.3--10~keV band for \xmm\ and \cxo\ spectra and in the 
3--30~keV band for \nustar\ spectra. All luminosities (Table S1) are in the 0.3--10~keV energy band 
and upper limits are at 3$\sigma$ confidence level. In the two \nustar\ and \xmm\ observations 
performed on 2013-11-06, we did not detect \srcfull\ down to a 3$\sigma$ limiting luminosity of 
2$\times$10$^{39}$ \lum\ and 3$\times$10$^{38}$ \lum, respectively, in agreement with the results 
by \cite{walton15}.  

Spectra are fit equally well by a broken power law (bknpow in XSPEC) or a Comptonized photons 
in a hot plasma model (comptt) or a multi-temperature disk blackbody with a power-law dependence 
of temperature on radius (diskpbb), all modified for the interstellar absorption, while simple 
thermal models or a power law does not fit adequately the data (reduced $\chi^2>2$). In our analysis 
we assumed chemical abundances from both  \cite{anders89} and \cite{wilms00} and cross sections from 
\cite{bcmc92},  providing consistent values for the single component parameters, with the latter 
abundances providing a higher, about 30\%, interstellar absorption value. 
For all models, the spectrum of \srcfull\ appears to change between the observations.

For the bknpow model, 
the best spectral fit, obtained by keeping the lower-energy photon index free,  resulted in a reduced 
$\chi^2$ of 1.07 (898 dof).
The best-fit parameters are $N_{\mathrm{H}} = (5.36\pm0.01) \times10^{21}$ cm$^{-2}$ for the absorption column, 
$E_{\mathrm{break}} = (6.7\pm0.2)$ keV for the energy of the break, $\Gamma_2 = 2.9\pm0.1$ for the power-law 
photon index above $E_{\mathrm{break}}$ and 
$\Gamma_1^{2003} = 1.58\pm0.02$, $\Gamma_1^{2012} = 1.28\pm0.03$, $\Gamma_1^{2013} = 1.91\pm0.03$, 
$\Gamma_1^{2014} = 1.53\pm0.02$ for the  power-law 
photon index below $E_{\mathrm{break}}$ for the four epochs (see also Table S1).

For the comptt model, 
the resulting best fit, obtained by keeping free the plasma temperature, has a reduced $\chi^2$ of 1.09 (898 dof).
The best-fit parameters are $N_{\mathrm{H}}  = (5.45\pm0.01) \times10^{21}$ cm$^{-2}$, Wien temperature of 
the input soft photons $kT_0 = (6.8\pm2.3) \times10^{-2}$ keV, optical depth $\tau = 7.69\pm0.01$, and 
plasma temperature $kT^{2003} = (2.5\pm0.1)$~keV, $kT^{2012} = (3.8\pm0.2)$ keV, $kT^{2013} = (2.0\pm0.2)$ 
keV, $kT^{2014} = (2.6\pm0.1)$ keV for the four epochs.

Finally, for the diskpbb model, 
the resulting best fit, assuming the disk temperature free to vary, has a reduced $\chi^2$ of 1.06 (898 dof).
The best-fit parameters are $N_{\mathrm{H}} = (4.86\pm0.01) \times10^{21}$ cm$^{-2}$, temperature at the 
inner disk radius $T_{in} = (3.4\pm0.1) \times10^{-2}$ keV and exponent of the radial dependence of the 
disk temperature $p^{2003} = (5.92\pm0.01) \times10^{-1}$, 
$p^{2012} = (6.62\pm0.01) \times10^{-1}$, $p^{2013} = (6.58\pm0.02) \times10^{-1}$, $p^{2014} = 
(6.03\pm0.01) \times10^{-1}$ for the four epochs.

The inferred 0.3-10\,keV luminosity measures are not strongly dependent on the models: in all cases the differences 
between the assumed models are within 1\%.
The  luminosities reported in Fig.\,S2 and Table S1  were instead the bolometric one inferred by adopting 
the bknpow model. 

We do not know why pulsations are detected only in two out of five \xmm\ data sets with sufficient time 
resolution, but we note that the spectral properties of the data (2003 and 2014) in which pulsations were 
detected  are similar to each other and different from the others. This is shown in Fig.\,S3, where we compare 
the spectra (fit with a bknpow model) collected in 2003, 2013 and 2014 (the \xmm\ and \nustar\ data 
obtained in 2013 are representative of the source spectral properties during which pulsations are not detected). 
For the two EPIC-pn data sets collected in 2003 and 2014, where pulsations are detected, we also carried out 
a pulse phase spectroscopy to study the possible presence of spectral variations as a function of the pulse 
phase. The results of the analysis confirms the presence of a phase shift of 0.15$\pm$0.03 cycles between 
soft ($<$5\,keV) and high ($>$7\,keV) energies in the 2014 EPIC-pn data (see Fig.\,S4.), while there is 
marginal significant dependence of the spectral parameters as a function of phase.  \\

\bigskip
\noindent {\bf \swift\ monitoring}

\noindent For each observation, we estimated the  source flux by fitting the 0.3--10~keV  spectrum with a 
broken power-law model (modified for the interstellar absorption), with the following (fixed) parameters: 
absorption $N_{\mathrm{H}}=5.3\times10^{21}$ cm$^{-2}$, break energy $E_{\mathrm{break}}=6.7$ keV, photon 
indices $\Gamma_1=1.6$ for $E<E_{\mathrm{break}}$ and $\Gamma_2=2.9$ for $E>E_{\mathrm{break}}$. 
Fig.\,S2. shows the long-term x-ray light curve of \srcfull\ (see also Table S1). The luminosity was 
computed over the 0.3--10~keV energy range from the unabsorbed flux assuming a distance of 17.1 Mpc; 
in those cases in which the count rate in an observation was compatible with zero, we set an upper limit 
at the 3$\sigma$ level. A $\sim$50\% modulation on a time scale of $\approx$80~d \cite{walton16} is 
apparent in the light curve obtained from 151 \swift\ observations from 2010 to 2016 (Fig.\,S2 and Table 
S1). The power spectrum peak at ${81\pm2}$~d has a ratio of centroid frequency to full width at half-maximum 
(a standard indicator for the coherence of a signal) of $\leq$50, amid typical values for quasi periodic 
oscillations and periodic signals, indicating that the modulation might not be strictly coherent. 
A super-orbital modulation of $\sim$55\,day  has been reported for M82~ULX-2 and interpreted as 
caused by a radiation-driven warping of the accretion disc \cite{kong16}. 
Considering that the M82~ULX-2 has an orbital period of 2.5\,day and assuming a linear scaling, the orbital 
period of \srcfull\ would be approximately 4\,day, which is similar to our estimate based on the \nustar\ 
2014 data (under the hypothesis that the two systems have a similar evolutionary history). \\

\noindent
{\Large\bf Supplementary text}

\noindent
{\large\bf The accretion models}

\noindent Accreting NSs release a luminosity of $L_{\rm acc} (R) = GM\dot M/R$
(where $M$ is the NS mass, $\dot M$ the mass accretion rate and $G$
the gravitational constant). $L_{\rm acc}$ may exceed $L_{\rm Edd}$ by
a large factor if the NS surface magnetic field is very high ($B >
10^{13}$~G), so that electron scattering cross sections in the
extraordinary mode are reduced for photon energies below the cyclotron
energy 
$E_c \sim 12 \ B_{12}$~keV (where $B_{12}=B/10^{12}$ {G}) 
and a high radiation flux can
escape without halting the accretion flow \cite{herold79}.
Detailed calculation show that magnetically-funneled column accretion onto
the NS poles can give rise to a luminosity of $\times 10^3\,{L_{\mathrm{Edd}}}$
if $B$ is $\sim 5 \times 10^{15}$~G \cite{mushtukov15}.
The solid line in Fig.\,3 shows the maximum luminosity that can be attained by
magnetically-funneled column accretion onto the NS poles 
as a function of the NS surface magnetic field [adopted from
Fig.5 of \cite{mushtukov15}]; above $\sim 10^{13}$~G it is well
approximated by $L_{39} \sim 0.35 \ B_{12}^{3/4}$, where 
$L_{39}=L_{\rm acc}/10^{39}~\mbox{erg}\,\mbox{s}^{-1}$.

Accretion at very-high rates is required to attain highly
super-Eddington luminosities; a number of conditions must be met for
it to take place.  First, the flow towards the magnetospheric
boundary, where the $B$-field begins to control the accreting plasma,
must be mediated by a disk, so that radiation emitted at the NS can
escape unimpeded in a range of directions.  For the magnetospheric
radius $r_{\mathrm{m}}$ we adopt the standard expression
\begin{equation}\tag{S6}
r_{\mathrm{m}}=\frac{\xi \mu^{4/7}}{\dot{M}^{2/7}(2GM)^{1/7}}=3.3\times10^{7}\,\xi_{0.5}\,B_{12}^{4/7}\,L_{39}^{-2/7}\,R_6^{10/7}\,m_{1.4}^{1/7}~\mbox{cm}
  \end{equation}
where $\mu=B R^3/2$ is the NS dipole magnetic moment, $B$ is the
magnetic field strength at the poles of the NS, $R$ is the NS radius,
and $\xi$ is a factor $<1$ expressing the magnetospheric radius as a
fraction of the Alfv\'en radius. The value on the right hand side was
evaluated by using $\xi_{0.5}=\xi/0.5$, 
$R_6=R/10^6~\mbox{cm}$ and $m_{1.4}=~M/1.4\,M_{\odot}$.

The accretion disk must remain
geometrically thin (with height/radius $<$1) down to the magnetospheric radius ($r_{\mathrm{m}}$,
where matter is channeled along the NS $B$-field lines), so as not to engulf the
magnetospheric boundary at high latitudes. This requires that the
accretion energy released in the disk is sub-Eddington, {\it i.e.}
$L_{\rm disk} (r_{\mathrm{m}}) = GM\dot M/2r_{\mathrm{m}} < L_{\rm Edd}$.
The corresponding threshold
\begin{equation}\tag{S7}
  L_{\rm acc}^{\rm max}=6.8\times10^{39}\,\xi_{0.5}^{7/9}\,B_{12}^{4/9}\,R_6^{1/3}\,m_{1.4}^{8/9}~ \mathrm{erg\,s}^{-1}
  \end{equation}
is plotted as a dashed line in Fig.\,3.

Another condition is that the angular velocity of the disk at $r_{\mathrm{m}}$
must be higher than the NS angular velocity, so that the centrifugal
drag exerted by the magnetic field lines as matter enters the
magnetosphere is weaker than gravity and matter can accrete onto the
NS surface.  This condition translates into $r_{\mathrm{m}} < r_{\mathrm{cor}} $, where
$r_{\mathrm{cor}} = \left(\frac{G M P^2}{4 \pi^2}\right)^{1/3}$ is the
corotation radius.  In the opposite case, $r_{\mathrm{m}} > r_{\mathrm{cor}}$, the centrifugal
drag by the magnetosphere at $r_{\mathrm{m}}$ exceeds gravity, accretion is
almost completely halted by the so-called propeller
mechanism and the NS enters a regime in which the emitted luminosity
is much lower \cite{illarionov75,stella86,tsygankov16,parfrey16,dallosso16}.
The minimum accretion luminosity below which the NS enters the propeller regime
is obtained by setting $r_{\mathrm{m}} = r_{\mathrm{cor}}$,
\begin{equation}\tag{S8}
  L_{\rm acc}^{\rm min}=2.8\times10^{36}\,\xi_{0.5}^{7/2}\,B_{12}^{2}\,R_6^{5}\,m_{1.4}^{-2/3}~ \mathrm{erg\,s}^{-1}
  \end{equation}
This is plotted as a dot-dashed line in Fig.\,3, for spin period of $P= 1.14$~s
 
If \srcfull\ emitted isotropically a maximum luminosity
$L_{\text{max,iso}} \sim 2 \times 10^{41}$~\lum, then a NS $B$-field at
the base of the accretion column of $\sim 5\times 10^{15}$~G
would be required \cite{mushtukov15}.  However, for such $B$ and a spin
period of $P\sim1$~s, the NS would be deep in the propeller regime and
accretion would not take place (see Fig.\,3).  On the contrary,
the NS in \srcfull\ was observed accreting over the luminosity range
from $L_{\text{max,iso}} \sim 2\times 10^{41}$~erg~s$^{-1}$ to 
$L_{\text{min,iso}} \sim L_{\text{max,iso}}/8$ 
(shown by the double-arrowed dashed segment
on the right of Fig.\,3), at a time-averaged luminosity of 
 $L_{\text{avg,iso}} \sim 10^{41}$~erg~s$^{-1}$
 (shown by the black circle on the double-arrowed segment in Fig.3).
  
 This inconsistency can be avoided if the source emission is beamed by
 a factor $b < 1$, so that its isotropic equivalent luminosity is
 $L_{\rm iso} = L_{\rm acc}/b$ and the accretion luminosity $L_{\rm
   acc} $ is reduced accordingly. We require that at the accretion
 luminosity corresponding to the minimum (detected) isotropic
 luminosity, {\it i.e.}  $L_{\rm min,acc} = L_{\rm min,iso} b$ the
 propeller mechanism has not yet set in, and that the surface
 $B$-field is such that the maximum accretion luminosity corresponding
 the maximum isotropic luminosity ($L_{\rm max,acc} = L_{\rm max,iso}
 b$) can be generated in the accreting column.  A beaming factor of
 $b\sim 1/100$ and a field of $B\sim 9\times10^{12}$~G satisfies both
 requirements, as shown in Fig.\,3 by the double-arrowed dashed
 segment on the left (note that in this case the requirement that the
 accretion disk remains geometrically thin at $r_{\mathrm{m}}$ does
 not provide any additional constraint).  One has to further verify
 that the mass accretion rate implied by $b\sim 1/100$ would be
 sufficient to cause the observed secular spin-up rate.  Accretion
 torques are highest when all the angular momentum of matter at
 $r_{\mathrm{m}} = r_{\mathrm{cor}}$ is transferred to the neutron
 star; this condition translates into an upper limit on the NS period
 derivative of $\dot P < \dot{M}\ r_{\mathrm{cor}}^2P/I$
(where $I \sim 10^{45}$~g~cm$^2$ is the NS moment of inertia). For the
(time-averaged) accretion rate of $\dot M\sim 5 \times 10^{18}$ g
s$^{-1}$ corresponding to $L_{\text{avg,iso}}$ for $b\sim 1/100$, the
above limit gives $\dot P <$- 2 $\times 10^{-10}$~s~s$^{-1}$ which is a
factor of $\sim$4 smaller than the secular value derived from the
data.  The latter is plotted as a double-dotted dashed line in Fig.3,
with the conversion factor between $L_{acc}$ and $\dot P$ shown by the
Y-axis scale on the right side.  Note also that the measured $\dot P$
might be affected by spin-down torques that set in during the
intervals in which the NS accreted at low levels or was in off-states,
possibly in the propeller regime.  Therefore we conclude that this
solution is untenable.

Implicit in standard accretion theory onto magnetic stars is the
assumption that the $B$-field is purely dipolar.  By analogy with
magnetars \cite{thompson95,tiengo13}, we then consider the possibility
that the $B$-field at the base of the accretion column is dominated by
higher multipoles rather than being a simple dipole [we note that this
  assumption does not necessarily imply the presence of a magnetar,
  the properties of which depend instead on the inner $B$-field
  \cite{rea10}; see also \cite{eksi15}] and that only the dipole
component survives at the magnetospheric radius ($r_{\mathrm{m}} \sim
10^8 $~cm~$\sim 100 R$ ), by virtue of its weaker radial
dependence. Such configuration retains a high value of the $B$-field
at the NS surface, thus permitting the release of super-Eddington
luminosities, while easing the propeller constraint at
$r_{\mathrm{m}}$. In practice one has to impose two conditions for the
dipole field at $r_{\mathrm{m}}$, namely that the accretion disk
remains thin for $L_{\rm max,acc} = L_{\rm max,iso} b $ and that the
propeller has not yet set in for $L_{\rm min,acc} = L_{\rm min,iso}
b$. A dipole field at the surface $B \sim 3 \times 10^{13}$~G and a
beaming of $b \sim 1/7$ satisfy both conditions (see the right
double-arrowed solid segment in Fig.\,3).

For the time-averaged accretion rate implied in the case $b \sim 1/7$
($\dot M\sim 7\times 10^{19}$ g s$^{-1}$), we derive a maximum spin-up
$\dot P$ of about $ -3 \times 10^{-9}$~s~s$^{-1}$, which accounts for
the measured secular value (this can also be seen in Fig.\,3: the
black circle representing the time-averaged accretion luminosity for
the $b = 1/7$ double-arrow lies above the double-dotted dashed line).
Somewhat smaller values of the beaming factor are also allowed, for
instance $b \sim 1/25$.  For such a value, the observed $\dot P$ could
still be produced (see the left double-arrowed solid segment in
Fig.\,3, whose black circle coincides with the double-dotted dashed
line); moreover the minimum accretion luminosity could attain a factor
of $\sim$4 value without yet triggering the propeller mechanism.  The
range of allowed solutions is represented by the grey-shaded
parallelogram in Fig.\,3 and comprises surface dipole fields between
$\sim$0.2 and $\sim 3 \times 10^{13}$~G.  Corresponding to that range,
a multipolar field $B_{\mathrm{multi}} > (0.7$--$3)\times 10^{14}$~G at
the NS surface (the reason why these should be regarded as lower
limits is explained below) is required to attain the maximum accretion
luminosity of $L_{\rm max,acc} =(0.8$--$3)\times 10^{40}$ \lum\ implied
by solutions with $b\sim 1/25$--$1/7$.  We thus conclude the
properties of \srcfull\ provide robust evidence in favor of a
multipolar $B$-field component at the star surface that is about an
order of magnitude larger than the dipole field component.

Finally, we remark that the model of \cite{mushtukov15}, which the
above discussion of the maximum luminosity of magnetic-column
accretion builds on, assumes a dipole magnetic field. Extending the
treatment to the case of higher order multipoles requires detailed
modeling. Yet some estimates can be made for a accretion columns
dominated by multipolar fields. An approximate value of the accretion
luminosity using eq. (8) of ref. \cite{mushtukov15}, is obtained by
retaining the dependence of the scattering cross section on $B$
($\sigma_\perp\sim B^{-2}$), which in turn changes with height inside
the column, $h$ (their expression (13) for the dipole; for higher
order multipoles the field decays faster with $h$). The maximum
luminosity $L_\mathrm{max}$ roughly corresponds to $H/R\sim 1$, where
$R$ is the star radius and $H$ is the total height of the emitting
column. We estimate that the maximum luminosity is a factor of a few
lower for higher multipole fields than for dipoles.  This is in line
with expectations, since the steeper decrease of multipolar fields
with height implies a larger cross section and hence a smaller
attainable luminosity. This is the reason why the value we inferred
above for $B_{\mathrm{multi}}$ should be regarded as lower limits. \\

\noindent
{\large\bf The nature of the system}

\noindent Before facing the possible nature of the binary system we briefly comment on the hypothesis that 
the source is not part of NGC\,5907. 
The possibility that \srcfull\ is a foreground high mass or low mass x-ray binary in our Galaxy is unlikely 
due to the high Galactic latitude and the Hubble Space Telescope (\hst) optical limit of 25 mag \cite{sutton13}. 

The observed spin-up rate of \srcfull\ and the accretion scenario imply an accretion
rate at least of $\dot{M}\sim10^{19}$~g~s$^{-1}$. A value this high rules out accretion through a wind.
Assuming Roche lobe overflow onto a NS and an orbital period of $\sim$ 4 days, stellar companions capable
to fill the Roche lobe are: (i) a star with $M_2 > 10$--15~$M_\odot$ at terminal age
main sequence or in the giant phase; (ii) a supergiant with a mass in excess of 10~$M_\odot$; (iii) a giant
with $M_2 \sim 1$~$M_\odot$; (iv) a giant with $M_2 = 2$--6~$M_\odot$.
We note that if the orbital period is longer than 4 days, a terminal age main sequence star would underfill 
its Roche lobe.
These high-mass binary system (HMXB, i and ii), low-mass binary system (LMXB,
iii), and intermediate-mass binary system (IMXB, iv) scenarios are consistent with the upper limits derived
from \hst\ images \cite{sutton13}. In the first
two cases, owing to the large mass ratio ($q=M_2/M_1\gg1$, with $M_1=1.4$~$M_\odot$), the mass transfer would
be unstable, unless the system loses mass at a very high rate
through powerful outflows and/or winds \cite{fragos15}. As an upper limit for the
evolutionary timescale, we can assume that the evolution proceeds on an approximately thermal timescale 
$t_{\mathrm{th}}$.
A rough estimate of the mass transfer rate can be obtained from ${\dot M_2} \approx M_2 / t_{\mathrm{th}}$.  
We
then have $t_{\mathrm{th}}\approx 7 \times 10^4$~yr and ${\dot{M}_2} \approx  10^{22}$~g~s$^{-1}$ for the 
scenario (i),
and $t_{\mathrm{th}}\approx3 \times 10^4$~yr and ${\dot{M}_2} \approx 2 \times 10^{22}$~g~s$^{-1}$ for (ii). The 
estimated upper limit on
${\dot{M}_2}$ is largely in excess of that required to power \srcfull\ . Thus, provided that the system 
can be stabilized 
through strong mass losses, these numbers are consistent with the possibility that \srcfull\ is an HMXB 
accreting
above the Eddington limit. On the other hand, if the outflows are not efficient, the system will enter the 
common envelope phase
and be very short lived.

In the LMXB scenario, the lower mass ratio allows stable mass transfer. During the core He burning phase,
the evolution likely proceeds on a nuclear timescale $t_{\mathrm{nuc}}$ while, during the H and He shell burning
phases, it proceeds on a thermal timescale.
Assuming that $\dot{M}_2 \approx M_2 / t_{\mathrm{nuc,th}}$, we have 
$t_{\mathrm{nuc}} \approx 7 \times 10^7$~yr and $\dot{M}_{2,\mathrm{nuc}} \approx 10^{19}$~g~s$^{-1}$ for the He core burning phase.
The orbital separation increases and exceeds rapidly a few tens of days \cite{tauris99},
which is difficult to reconcile with the interval estimated from our likelihood analysis.
The inferred mass transfer rate is sufficiently high to guarantee a stable accretion disc
\cite{dubus99}, but follows a bit short of what is needed to power \srcfull\ even accounting for beaming.
In addition, we note that LMXB ULXs seem to be preferentially associated with elliptical galaxies and their
average luminosity is lower than that of ULXs in spiral galaxies \cite{swartz11}. The fraction of ULXs
associated with old stellar populations in spiral galaxies is rather low\cite{liu06}, implying that ULX
LMXB systems possibly similar to \srcfull\ might not be common.

The last possibility is that of an intermediate mass donor. Post-main sequence stars of $\sim$2--6~$M_\odot$ 
without deep 
convective envelopes are able to survive a highly super-Eddington mass transfer phase and remain dynamically
stable for a few hundred thousands years, while maintaining orbital periods of a few days \cite{tauris00}.
This IMBH channel appears to be consistent with the inferred properties of \srcfull. \\

\clearpage

\begin{figure}
\resizebox{\hsize}{!}{\includegraphics{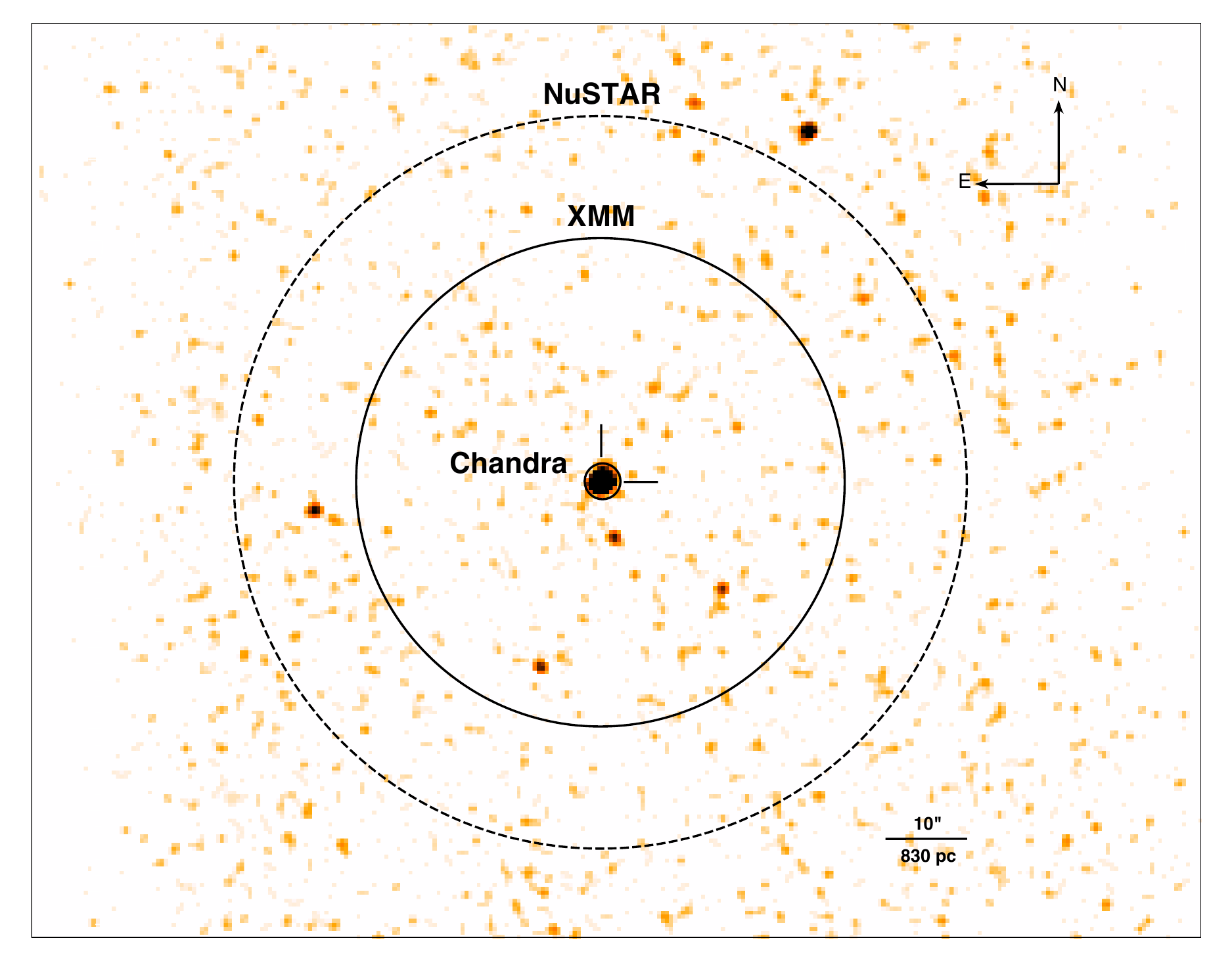}}
\noindent {\bf Fig. S1. Chandra image of the region around NGC 5907 ULX.} ~ \cxo\ ACIS-S 0.2--10~keV  (observation 
identification number: 12987) image of the $120''\times120''$ region around 
\srcfull\ with superimposed the \xmm\ (solid line) and \nustar\ (stepped line) regions used in 
this work to extract the source events. In order to emphasize the sources the \cxo\ imaged has been 
smoothed by using a 2$''$-radius Gaussian function. At the distance of 17.1\,Mpc, an angular separation of 10$''$ 
correspond to about 830\,pc.
\end{figure}
\clearpage

\clearpage
\begin{figure}
\resizebox{\hsize}{!}{\includegraphics{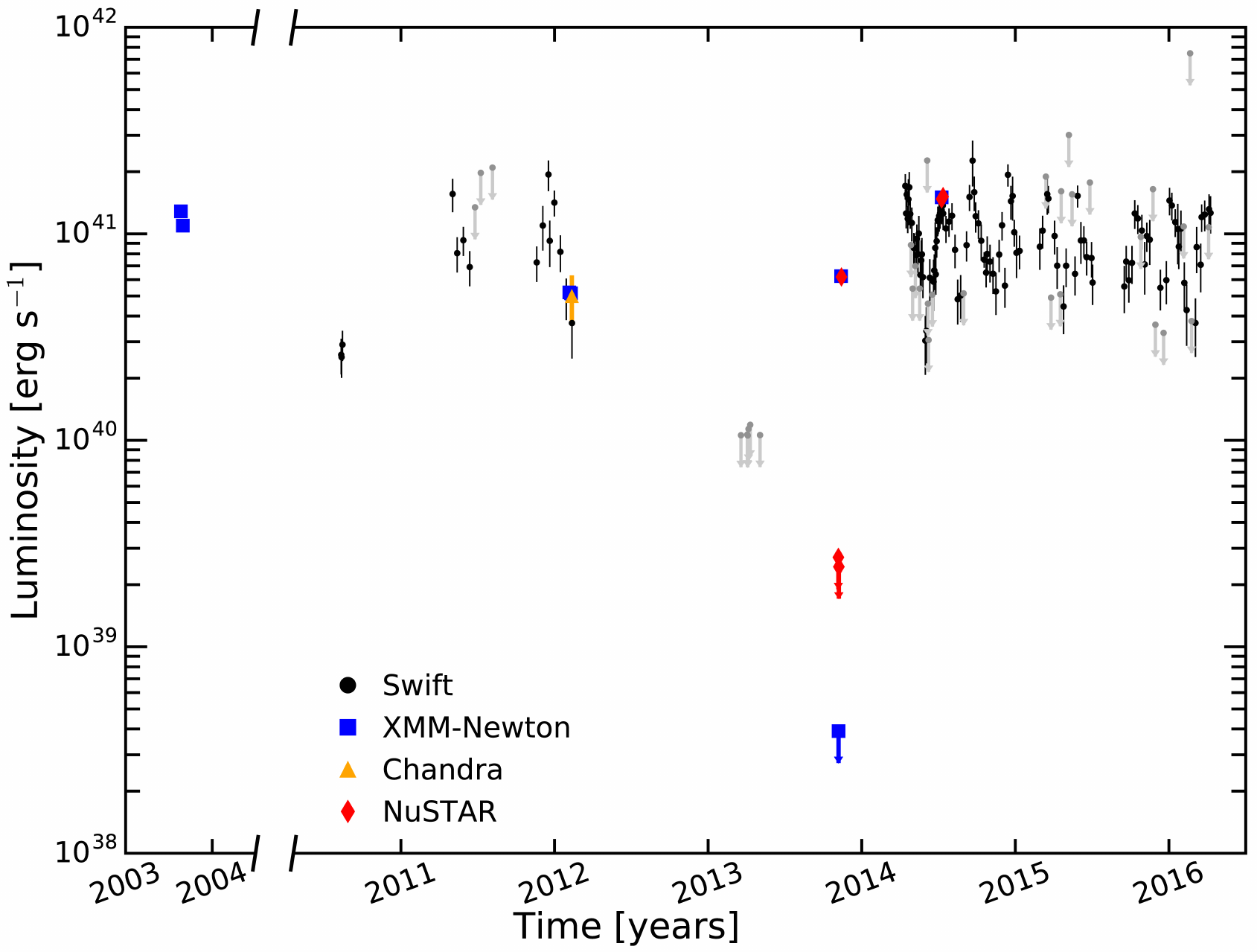}}
\noindent {\bf Fig. S2. Long-term light curve of NGC\,5907 ULX.} ~ Long-term multi-mission light curve of \srcfull. 
The luminosity was computed 
assuming a distance of 17.1~Mpc. All errors are at 1$\sigma$ confidence level, while upper limits 
(red, blue and grey down arrows for \nustar, \xmm\ and \swift, respectively) are at 3$\sigma$ level.
\end{figure}
\clearpage 

\begin{figure}
\resizebox{\hsize}{!}{\includegraphics[angle=-90]{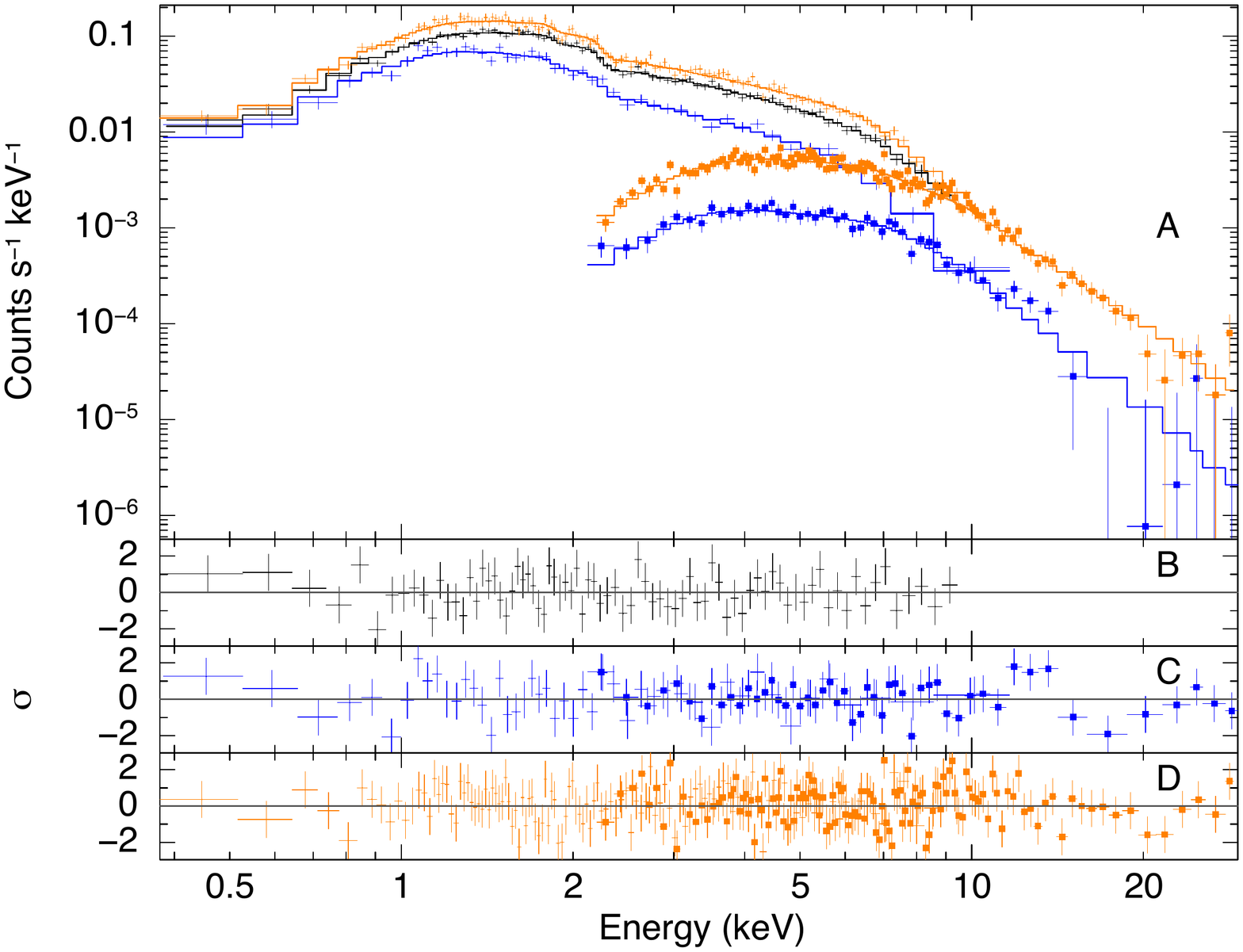}}\vspace{.2cm}
{\bf Fig. S3. Spectral energy distribution of NGC\,5907 ULX.} ~ \xmm\ EPIC pn (crosses) and \nustar\ FPM 
(squares) spectra collected in 2003 (black, only pn), 
2013 (blue), and 2014 (orange; panel A). For the two latest epochs, the \xmm\ and \nustar\ data are 
simultaneous. The solid lines show the broken power-law model. Lower panel: The residuals (in units of 
standard deviation) are reported separately for the three epochs: 2003 (panel B), 2013 (panel C) and 2014 
(panel D).  
\end{figure}
\clearpage
 
\begin{figure}
\resizebox{\hsize}{!}{\includegraphics{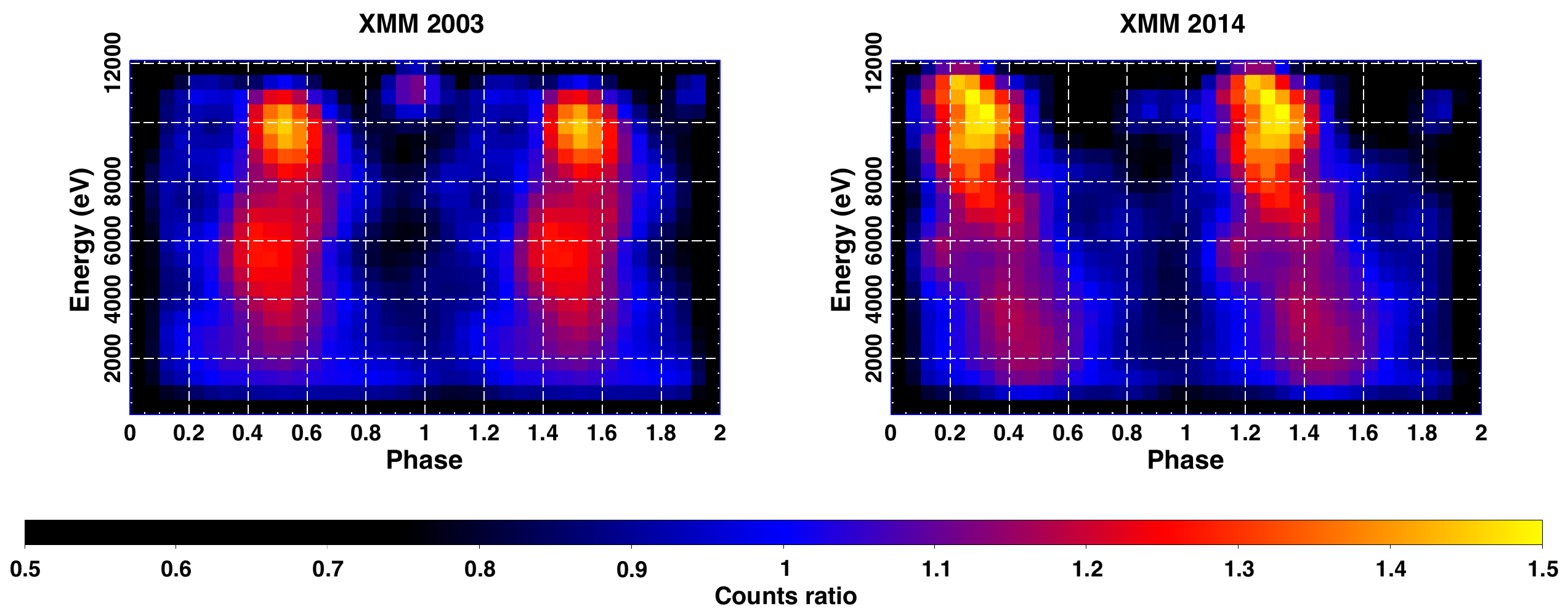}}
{\bf Fig. S4. Energy--phase distribution of NGC\,5907 ULX.} ~ Phase-energy images obtained by binning the 
EPIC pn source counts into
20 phase bins and energy channels of 500 eV for the 2003 (left panel) and 2014 (right panel) \xmm\  
observations. The color scale represents the ratio between the counts in each phase interval and the 
phase-averaged counts in the same energy bin. A pulse peak shift  is evident in the 
2014 dataset.
\end{figure}

\begin{table}
  \noindent {{\bf Table\,S1.  X-ray observations log}. The bolometric luminosity is inferred by assuming 
  a distance of 17.1 Mpc and extrapolating the result of the fit with a broken power-law model. 
  Uncertainties are at 1$\sigma$ confidence level. Upper limits are at 3$\sigma$ level. The symbol ($^{\ast}$) marks 
  the observations during which pulsations were detected. The full table is available in a separate text file 
  as part of the supplementary material.}\label{tab}
\begin{footnotesize}
\begin{tabular}{clccc}
\hline
\hline
{\bf Observation id.} & {\bf Mission/Instr.}& {\bf Start date} & {\bf Net exposure} & {\bf Luminosity}\\
 & &{\bf (yyyy-mm-dd)} & {\bf (s)} & {\bf ($10^{40}$ erg s$^{-1}$)}\\
\hline
\phantom{$^{\ast}$}0145190201$^{\ast}$ & \xmm/EPIC & 2003-02-20  & 36898 & $12.9\pm0.1$\\
0145190101 & \xmm/EPIC & 2003-02-28 & 42095 & $11.0\pm0.1$\\
00031785001 & \swift/XRT & 2010-08-12  & 5516 & $2.6\pm0.3$\\
00031785002 & \swift/XRT & 2010-08-13 & 5151 & $2.5\pm0.3$\\
00031785003 & \swift/XRT & 2010-08-15 & 6710 & $2.9\pm0.3$\\
00031785005 & \swift/XRT & 2011-05-04    & 1178 & $15.7\pm1.9$\\
00031785006 & \swift/XRT & 2011-05-15  & 1905 & $8\pm1$\\
00031785007 & \swift/XRT & 2011-05-29   & 2552 & $9\pm1$\\
00031785008 & \swift/XRT & 2011-06-13    & 2102 & $7.0\pm0.9$\\
00031785009 & \swift/XRT & 2011-06-26   & 254 & $<$13\\
00031785010 & \swift/XRT & 2011-07-10    & 172 & $<$20\\
00031785011 & \swift/XRT & 2011-08-07  & 162 & $<$22\\
00031785012 & \swift/XRT & 2011-11-20   & 2040 & $7.3\pm0.9$\\
00031785013 & \swift/XRT & 2011-12-04   & 899 & $11\pm2$\\
00031785014 & \swift/XRT & 2011-12-18    & 951 & $20\pm2$\\
00031785015 & \swift/XRT & 2011-12-21    & 916 & $9\pm1$\\
00031785016 & \swift/XRT & 2012-01-01    & 1860 & $14\pm1$\\
00031785017 & \swift/XRT & 2012-01-15    & 1655 & $8\pm1$\\
00031785018 & \swift/XRT & 2012-01-29   & 2157 & $5.0\pm0.7$\\
0673920201 & \xmm/EPIC & 2012-02-05 & 14993 & $5.3\pm0.1$\\
0673920301 & \xmm/EPIC & 2012-02-09 & 15024 & $5.1\pm0.1$\\
12987 & \cxo/ACIS-S & 2012-02-11& 15977 & $5.0\pm0.8$\\
14391 & \cxo/ACIS-S & 2012-02-11 & 13086 & $5.0\pm0.8$\\
00031785019 & \swift/XRT & 2012-02-12   & 1445 & $3.7\pm0.8$\\
%
%
00032764001 & \swift/XRT & 2013-03-19    & 3958 & $<$1\\
00032764002 & \swift/XRT & 2013-04-03    & 3925 & $<$1\\
00032764003 & \swift/XRT & 2013-04-04    & 3970 & $<$2\\
00032764004 & \swift/XRT & 2013-04-06   & 3643 & $<$1\\
00032764005 & \swift/XRT & 2013-04-10  & 3451 & $<$1\\
00032764006 & \swift/XRT & 2013-05-04  & 3950 & $<$1\\
0724810201 & \xmm/EPIC & 2013-11-06 & 30341 & $<$0.04\\
30002039002 & \nustar/FPM & 2013-11-06 & 45339 & $<$0.3\\
30002039003 & \nustar/FPM & 2013-11-06& 68542 & $<$0.3\\
0724810401 & \xmm/EPIC & 2013-11-12 & 29961 & 6.2$\pm$0.1\\
30002039005 & \nustar/FPM & 2013-11-12 & 112924 & 6.2$\pm$0.1\\
00032764007 & \swift/XRT & 2014-04-14   & 1828 & 17$\pm$2\\
00032764008 & \swift/XRT & 2014-04-16   & 1927 & 12$\pm$1\\
00032764009 & \swift/XRT & 2014-04-18   & 2245 & 16$\pm$1\\
00032764010 & \swift/XRT & 2014-04-20   & 2160 & 12$\pm$1\\
00032764011 & \swift/XRT & 2014-04-22    & 914 & 14$\pm$2\\
00032764012 & \swift/XRT & 2014-04-24  & 938 & 17$\pm$2\\
00032764013 & \swift/XRT & 2014-04-26   & 1935 & 12$\pm$1\\
00032764014 & \swift/XRT & 2014-04-28   & 392 & 23$\pm$3\\
\end{tabular}
\end{footnotesize}
\end{table}

\begin{table}
\begin{center}
Table\,S1 --- Continued
\end{center}
\begin{footnotesize}
\begin{tabular}{clccc}
\hline
{\bf Observation id.} & {\bf Mission/Instr.}& {\bf Start/Stop time}  & {\bf Net exposure} & {\bf Luminosity$^{a}$}\\
 & {\bf (yyyy-mm-dd)} & {\bf (s)} & {\bf ($10^{40}$ erg s$^{-1}$)}\\
\hline
00032764015 & \swift/XRT & 2014-04-30  & 1583 & 11$\pm$1\\
00032764016 & \swift/XRT & 2014-05-02   & 649 & $<$6\\
00032764017 & \swift/XRT & 2014-05-04   & 1972 & 9$\pm$1\\
00032764018 & \swift/XRT & 2014-05-06    & 2042 & 9$\pm$1\\
00032764019 & \swift/XRT & 2014-05-08    & 499 & $<$8\\
00032764020 & \swift/XRT & 2014-05-10    & 1900 & 8$\pm$1\\
00032764021 & \swift/XRT & 2014-05-12   & 1927 & 12$\pm$1\\
00032764022 & \swift/XRT & 2014-05-14    & 1937 & 8$\pm$1\\
00032764023 & \swift/XRT & 2014-05-16    & 1485 & 9$\pm$1\\
00032764024 & \swift/XRT & 2014-05-18   & 649 & $<$6\\
00032764025 & \swift/XRT & 2014-05-20    & 1688 & 6$\pm$1\\
00032764026 & \swift/XRT & 2014-05-22    & 2067 & 8$\pm$1\\
00032764027 & \swift/XRT & 2014-05-24    & 1897 & 8$\pm$1\\
00032764028 & \swift/XRT & 2014-05-26    & 2075 & 8$\pm$1\\
00032764029 & \swift/XRT & 2014-06-01  & 2015 & 3.1$\pm$0.6\\
00032764030 & \swift/XRT & 2014-06-03   & 1853 & 3.4$\pm$0.7\\
00032764031 & \swift/XRT & 2014-06-05   & 149 & $<$23\\
00032764032 & \swift/XRT & 2014-06-07   & 774 & $<$5\\
00032764033 & \swift/XRT & 2014-06-09  & 1193 & $<$3\\
00032764034 & \swift/XRT & 2014-06-11   & 1673 & 6$\pm$1\\
00032764035 & \swift/XRT & 2014-06-18   & 696 & $<$5\\
00032764036 & \swift/XRT & 2014-06-20   & 1952 & 6$\pm$1\\
00032764037 &\swift/XRT &  2014-06-22  & 2082 & 6$\pm$1\\
00032764038 &\swift/XRT &  2014-06-24    & 586 & 9$\pm$2\\
00032764039 &\swift/XRT &  2014-06-26   & 2065 & 6.4$\pm$0.8\\
00032764040 & \swift/XRT & 2014-06-28 & 2097 & 9$\pm$1\\
00032764041 & \swift/XRT & 2014-06-30    & 1790 & 12$\pm$1\\
00032764042 & \swift/XRT & 2014-07-02  & 2609 & 11$\pm$1\\
00032764043 & \swift/XRT & 2014-07-04  & 1840 & 12$\pm$1\\
00032764044 & \swift/XRT & 2014-07-06   & 1887 & 12$\pm$1\\
0729561301$^b$ & \xmm/EPIC & 2014-07-09 & 37569 & 15.0$\pm$0.1\\
80001042002$^b$  & \nustar/FPM & 2014-07-09 & 57113 & 14.7$\pm$0.2\\
00080756001 & \swift/XRT & 2014-07-10   & 1937 & 8$\pm$1\\
00080756002 & \swift/XRT & 2014-07-12    & 1872 & 14$\pm$1\\
80001042004$^b$  & \nustar/FPM & 2014-07-12 & 56312 & 15.2$\pm$0.2\\
00032764045 & \swift/XRT & 2014-07-13   & 1600 & 12$\pm$1\\
00032764046 & \swift/XRT & 2014-07-20   & 2290 & 11$\pm$1\\
00032764047 &\swift/XRT &  2014-07-27    & 1947 & 11$\pm$1\\
00032764048 & \swift/XRT & 2014-08-03   & 1945 & 12$\pm$1\\
00032764049 & \swift/XRT & 2014-08-10    & 1955 & 12$\pm$1\\
00032764050 & \swift/XRT & 2014-08-17    & 1912 & 4.8$\pm$0.8\\
00032764051 & \swift/XRT & 2014-08-24    & 1658 & 5.1$\pm$0.8\\
00032764052 & \swift/XRT & 2014-08-31    & 686 & $<$5\\
00032764053 & \swift/XRT & 2014-09-07    & 2205 & 11$\pm$1\\
\end{tabular}
\end{footnotesize}
\end{table}

\begin{table}
\begin{center}
Table\,S1 --- Continued
\end{center}
\begin{footnotesize}
\begin{tabular}{clccc}
\hline
{\bf Observation id.} &{\bf Mission/Instr.}&  {\bf Start/Stop time}  & {\bf Net exposure} & {\bf Luminosity$^{a}$}\\
 &{\bf (yyyy-mm-dd)} & {\bf (s)} & {\bf ($10^{40}$ erg s$^{-1}$)}\\
\hline
00032764054 & \swift/XRT & 2014-09-14    & 1723 & 14$\pm$1\\
00032764055 & \swift/XRT & 2014-09-21    & 392 & 23$\pm$3\\
00032764056 & \swift/XRT & 2014-09-25    & 1031 & 16$\pm$2\\
00032764057 & \swift/XRT & 2014-09-28    & 1585 & 12$\pm$1\\
00032764058 & \swift/XRT & 2014-10-05    & 1975 & 11$\pm$1\\
00032764059 & \swift/XRT & 2014-10-12    & 1902 & 9$\pm$1\\
00032764060 & \swift/XRT & 2014-10-19    & 9589 & 7.4$\pm$0.4\\
00032764061 & \swift/XRT & 2014-10-23    & 3835 & 6.5$\pm$0.6\\
00032764062 & \swift/XRT & 2014-10-26    & 1445 & 3.7$\pm$0.8\\
00032764063 & \swift/XRT & 2014-11-02    & 1748 & 8$\pm$1\\
00032764064 & \swift/XRT & 2014-11-09    & 2195 & 6.4$\pm$0.8\\
00032764065 & \swift/XRT & 2014-11-16    & 2270 & 5.3$\pm$0.8\\
00032764066 & \swift/XRT & 2014-11-23    & 2232 & 8$\pm$1\\
00032764067 & \swift/XRT & 2014-11-30    & 1920 & 11$\pm$1\\
00032764068 & \swift/XRT & 2014-12-07    & 2000 & 5.6$\pm$0.8\\
00032764069 & \swift/XRT & 2014-12-14    & 2012 & 19$\pm$2\\
00032764070 & \swift/XRT & 2014-12-21    & 1046 & 14$\pm$2\\
00032764071 & \swift/XRT & 2014-12-25    & 953 & 15$\pm$2\\
00032764072 & \swift/XRT & 2014-12-29    & 2427 & 11$\pm$1\\
00032764073 & \swift/XRT & 2015-01-04    & 1383 & 8$\pm$1\\
00032764074 & \swift/XRT & 2015-01-11    & 1970 & 8$\pm$1\\
00032764075 & \swift/XRT & 2015-02-28    & 1430 & 9$\pm$1\\
00032764076 & \swift/XRT & 2015-03-07    & 1680 & 11$\pm$1\\
00032764077 & \swift/XRT & 2015-03-14    & 179 & $<$19\\
00032764078 & \swift/XRT & 2015-03-18   & 816 & 10$\pm$3\\
00032764079 & \swift/XRT & 2015-03-20    & 1910 & 16$\pm$1\\
00032764080 & \swift/XRT & 2015-03-27    & 721 & $<$5\\
00032764081 & \swift/XRT & 2015-04-04  & 1573 & 9$\pm$1\\
00032764082 & \swift/XRT & 2015-04-10    & 1555 & 8$\pm$1\\
00032764083 & \swift/XRT & 2015-04-18    & 694 & $<$5\\
00032764084 & \swift/XRT & 2015-04-20    & 212 & $<$16\\
00032764085 & \swift/XRT & 2015-04-26    & 1932 & 4.5$\pm$0.8\\
00032764086 & \swift/XRT & 2015-05-02    & 1805 & 8$\pm$1\\
00032764087 & \swift/XRT & 2015-05-08    & 112 & $<$29\\
00032764088 & \swift/XRT & 2015-05-16    & 219 & $<$16\\
00032764089 & \swift/XRT & 2015-05-23    & 2262 & 4$\pm$1\\
00032764090 & \swift/XRT & 2015-05-29    & 2459 & 16$\pm$1\\
00032764091 & \swift/XRT & 2015-06-06    & 1206 & 9$\pm$1\\
00032764092 & \swift/XRT & 2015-06-13    & 1975 & 11$\pm$1\\
00032764093 & \swift/XRT & 2015-06-19    & 2325 & 8$\pm$1\\
00032764094 & \swift/XRT & 2015-06-27    & 192 & $<$17\\
00032764095 & \swift/XRT & 2015-07-01    & 1853 & 3.4$\pm$0.7\\
00032764096 & \swift/XRT & 2015-07-04    & 1915 & 5.9$\pm$0.8\\
00032764097 & \swift/XRT & 2015-09-17    & 1480 & 5.6$\pm$0.9\\

\end{tabular}
\end{footnotesize}
\end{table}

\begin{table}
\begin{center}
Table\,S1 --- Continued
\end{center}
\begin{footnotesize}
\begin{tabular}{clccc}
\hline
{\bf Observation id.} & {\bf Mission/Instr.}& {\bf Start/Stop time}  & {\bf Net exposure} & {\bf Luminosity$^{a}$}\\
 &{\bf (yyyy-mm-dd)} & {\bf (s)} & {\bf ($10^{40}$ erg s$^{-1}$)}\\
\hline
00032764098 & \swift/XRT & 2015-09-21    & 2057 & $8\pm1$\\
00032764099 & \swift/XRT & 2015-09-28    & 1905 & $8\pm1$\\
00032764100 & \swift/XRT & 2015-10-05    & 2030 & $8\pm1$\\
00032764101 & \swift/XRT & 2015-10-12    & 1733 & $12\pm1$\\
00032764102 & \swift/XRT & 2015-10-19    & 1810 & $12\pm1$\\
00032764103 & \swift/XRT & 2015-10-26    & 357 & $<$9\\
00032764104 & \swift/XRT & 2015-10-29    & 1448 & 11$\pm$1\\
00032764105 & \swift/XRT & 2015-11-04    & 926 & 8$\pm$1\\
00032764106 & \swift/XRT & 2015-11-09    & 2135 & 9$\pm$1\\
00032764107 & \swift/XRT & 2015-11-16    & 931 & 9$\pm$1\\
00032764109 & \swift/XRT & 2015-11-24    & 207 & $<$17\\
00032764110 & \swift/XRT & 2015-11-30    & 993 & $<$3\\
00032764111 & \swift/XRT & 2015-12-12    & 2002 & 5.5$\pm$0.8\\
00032764112 & \swift/XRT & 2015-12-19    & 1096 & $<$3\\
00032764113 & \swift/XRT & 2015-12-26    & 1808 & 6$\pm$1\\
00032764114 & \swift/XRT & 2016-01-02    & 1613 & 14$\pm$1\\
00032764115 &\swift/XRT &  2016-01-08    & 1620 & 14$\pm$1\\
00032764116 & \swift/XRT & 2016-01-16    & 1967 & 11$\pm$1\\
00032764117 & \swift/XRT & 2016-01-22    & 496 & 11$\pm$2\\
00032764118 & \swift/XRT & 2016-01-24    & 1480 & 5.6$\pm$0.9\\
00032764119 & \swift/XRT & 2016-01-30    & 1668 & 11$\pm$2\\
00032764120 & \swift/XRT & 2016-02-05    & 317 & $<$11\\
00032764121 & \swift/XRT & 2016-02-07    & 1410 & 6$\pm$1\\
00032764122 & \swift/XRT & 2016-02-12   & 1216 & 4.3$\pm$0.9\\
00032764123 & \swift/XRT & 2016-02-20    & 44 & $<$74\\
00032764124 &\swift/XRT &  2016-02-24   & 951 & 20$\pm$2\\
00032764126 & \swift/XRT & 2016-03-04    & 1648 & 3.7$\pm$0.8\\
00032764127 & \swift/XRT & 2016-03-07    & 1675 & 9$\pm$1\\
00032764129 & \swift/XRT & 2016-03-17   & 1246 & 8$\pm$1\\
00032764130 &\swift/XRT &  2016-03-19    & 1947 & 12$\pm$1\\
00032764131 & \swift/XRT & 2016-03-26    & 1550 & 12$\pm$1\\
00032764132 & \swift/XRT & 2016-04-04    & 319 & $<$11\\
00032764133 & \swift/XRT & 2016-04-05    & 1278 & 12$\pm$2\\
00032764134 & \swift/XRT & 2016-04-09    & 1111 & 12$\pm$2\\
\hline \\
\end{tabular}

$^a$ {The isotropic bolometric luminosity is inferred by assuming a distance of 17.1\,Mpc. 
  Uncertainties are at 1$\sigma$ level and upper limits at 3$\sigma$ level in the same band.}\\
$^b$ {Observations during which pulsations were detected.}

\end{footnotesize}
\end{table}

\end{document}